\magnification=\magstep1
\tolerance=500
\rightline{19 February, 2020}
\bigskip
\centerline{\bf SYMMETRY OF THE RELATIVISTIC TWO BODY BOUND STATE}
\bigskip
\centerline{ L.P. Horwitz$^{1,2,3}$ and R.I. Arshansky$^4$\footnote{*}{This article is based on the work of Arshansky and Horwitz[18]}}

\bigskip
\centerline{${}^1$ School of Physics, Tel Aviv University, Ramat Aviv
69978, Israel}
\centerline{${}^2$ Department of Physics, Bar Ilan University, Ramat Gan
52900, Israel}
\centerline{${}^3$ Department of Physics, Ariel University, Ariel 40700, Israel}
\centerline{${}^4$ Givat Zorfatit Etzel 12/14 Jerusalem 9754}
\bigskip
\noindent Received: 19 January 2020; Accepted: 10 Februry 2020; Published date
\bigskip
\noindent email:larry@tauex.tau.ac.il
\bigskip
\noindent{\it Abstract}
\smallskip
\par We show that in a relativistically covariant formulation of the two body problem, the bound state spectrum is in agreement, up to relativistic corrections, with the nonrelativistic bound state spectrum. This solution is achieved by solving the problem with support of the wave functions in an $O(2,1)$ invariant submanifold of the Minkowski spacetime. The $O(3,1)$ invariance of the differential equation requires, however, that the solutions provide a representation of $O(3,1)$. Such solutions are obtained by means of the method of induced representations, providing a basic insight into the subject of the symmetries of relativistic dynamics.
\bigskip
\noindent{\bf 1. Introduction}
\bigskip
\par  In the nonrelativistic Newtonian-Galilean view,  two particles may be
thought of as interacting through a potential function $V({\bf x}_1(t),
{\bf x}_2(t))$; for Galiliean invariance, $V$ must be a scalar function
of the difference, {\it i.e.},$V({\bf x}_1(t)-{\bf x}_2(t))$. In such a potential model, ${\bf x}_1$ and ${\bf x}_2$ are taken to be at
equal time, corresponding to a correlation between the two particles
consistent with the Newtonian-Galilean picture.
\par For the relativistic theory, two  world lines
with action at a distance interaction between two points $x_1^\mu$ and $x_2^\mu$ cannot be  correlated by the variable $t$ in every frame.
\par The Stueckelberg (SHP) theory[1] provides an effective and systematic way of dealing with the $N$ body problem, and has been
applied in describing relativistic fluid mechanics [2],
 the Gibbs ensembles in statistical mechanics and the
Boltzmann equation [3], systems of many
identical particles [4], and other
applications. 
\par The basic idea of the SHP theory is the parametrization of the world lines of particles with a universal parameter $\tau$ [5](see also [6][7]). Stueckelberg [8] described {\it classical} pair annihilation with a world line that proceeds, in $\tau$, in the positive direction of the time $t$ (the observable time of Einstein [9]) and then passes to a motion in the negative direction of time for $\tau$ proceeding in its monotonic development, precisely as postulated by Newton [10][11]. The transition is caused by interaction, such as emission of a photon. Although this process was considered to be classical, it occurs in a diagram in Feynman's perturbative expansion of the $S$-matrix [12].
\par Stueckelberg [8] then considered the symplectic manifold of $\{ x^\mu, p_\mu\}$, with $\mu,\nu = (0,1,2,3)$ with diagonal metric $\eta_{\mu\nu} = (-,+,+,+)$ (raising and lowering indices).  Here, $x^\mu= \{x^0, x^1,x^2,x^3\}$, where $x^0 = ct\footnote{*}{ For $c\rightarrow \infty$, $ct$ may remain finite for $t\rightarrow 0$ and can be taken to be an arbitrary constant. }, p^0= E/c$. We shall generally write $c=1$ but note that in the nonrelativisic (NR) limit, $c \rightarrow \infty$, so that $p^0 \rightarrow 0$ for Pfinite energy $E$. Stueckelberg then wrote an invariant Hamiltonian of the form (for $V(x)$ scalar)
$$ K = {p^\mu p_\mu \over 2M} + V(x), \eqno(1)$$
which goes over to the usual NR Hamiltonian for in the NR limit.
\par He assumed the equations of motion
$$ \eqalign{{\dot x}^\mu\equiv {dx^\mu \over d\tau}  &= {\partial K \over \partial p_\mu}\cr
{\dot p}_\mu &= -{\partial K \over \partial x^\mu}\equiv {dp_\mu \over d\tau}\cr} \eqno(2)$$
It then follows from $(1)$ that the proper time $ds^2= -dx^\mu dx_\mu $ satisfies
$$ {ds^2 \over d\tau^2} =- {p^\mu p_\mu \over M^2} = {m^2 \over M^2}. \eqno(3)$$
The theory implies that the particle mass $m$ is a dynamical variable, reflecting the fact that the Einstein time $t$ is an observable, and therefore that $E= \pm \sqrt{{\bf p}^2 + m^2}$, conjugate  to $t$,  must be an observable as well [5]. For $m^2=  M^2$, $(3)$ implies that the square of the proper time interval is equal to $(d\tau)^2$, but in general, this relation cannot be maintained for non-trivial interaction.
\par The Poisson bracket structure then follows from $(2)$. The $\tau$ derivative of a function of $x,p$ is given by
$$ \eqalign{{d\over d\tau} F(x,p) &= {\partial F\over \partial x^\mu} {dx^\mu \over d\tau} + {\partial F \over \partial p_\mu} {dp_\mu \over d\tau} \cr
&= {\partial F\over \partial x^\mu}{\partial K \over \partial p_\mu} -{\partial F \over \partial p_\mu}{\partial K \over \partial x^\mu}\cr
&\equiv [F,K]_{PB}; \cr} \eqno(4)$$
With this,we see that
$$ [x^\mu, p_\nu]_{PB}= {\delta^\mu}_\nu. \eqno(5)$$
Following Dirac [13], it is assumed that the operator commutation relations, following the group action of translation implied by the Poisson bracket,
$$[x^\mu, p_\nu]= i\hbar {\delta^\mu}_\nu \eqno(6)$$
as the basis for the construction Pof the quantum theory [5].
\par The correspondig Stueckelberg-Schr\"odinger equation is then taken to be, derived from the unitary evolution of the wave function $\psi_\tau(x)$,
$$ i \hbar{\partial \psi_\tau(x) \over \partial \tau}= K\psi_\tau(x), \eqno(7)$$
with the operators $p_\mu$ in $K$ represented as $-i\hbar {\partial\over\partial x^\mu}$, self-adjoint in the scalar product $(\psi,\chi) =\int d^4x {\psi_\tau}^*(x) {\chi_\tau}(x)$.
\par Eq. $(7)$ corresponds to the quantum one particle problem. We now proceed to discuss the two body problem.

\bigskip
\noindent{\bf 2. The two body bounPPPd state.}
\bigskip
\par We review here the relativistic two body problem with
invariant action at a distance potentials, for bound states. 
\par As a candidate for an invariant action at a distance potential
for the two body relativistic bound state we take for the potential $V$ the function $V(\rho)$, for
$$\rho^2 = ({\bf x}_1 - {\bf x}_2)^2 - (t_1 -t_2)^2 \equiv {\bf x}^2 -
t^2 ,\eqno(8)$$
where $x^\mu_1$ and $x^\mu_2$ are taken at equal $\tau$, acting as a
correlation parameter as well as the global generating parameter of evolution. 
This ``relative coordinate'' (squared) reduces to  $({\bf x}_1 - {\bf x}_2)^2 
\equiv {\bf x}^2$ at equal time for the two particles in the
nonrelativistic limit, so that $ \rho$ becomes $r$ in this limit (for simultaneous $t_1$ and $t_2$). Clearly, the solutions of a problem
with this potential must then reduce to the solutions of the
corresponding nonrelativistic problem in that limit.
\par The two body Stueckelberg Hamiltonian, is 
$$ K = {{p_1}^\mu {p_1}_\mu \over 2M_1} +  {{p_2}^\mu {p_2}_\mu \over
2M_2} + V(x). \eqno(9)$$
Since $K$ does not depend on the total
(spacetime) ``center of mass'' P
$$ X^\mu = {M_1 x_1^\mu + M_2 x_2^\mu \over M_1 +M_2}, \eqno(10)$$
the two body Hamiltonian can be separated into the sum of two
Hamiltonians, one
for the ``center of mass'' motion and the second for the relative
motion, by defining the total momentum, which is absolutely conserved,
$$ P^\mu = p_1^\mu + p_2^\mu \eqno(11PP)$$
and the relative motion momentum
$$ p^\mu = {M_2 p_1^\mu- M_1 p_2^\mu \over M_1 + M_2}. \eqno(12)$$
The pairs $P^\mu, X^\mu$ and
$p^\mu,x^\mu$ satisfy separately the canonical Poisson bracket
(classically) and commutation relations (quantum mechanically), and
commute with each other. 
Then 
$$\eqalign{ K &= {P^\mu P_\mu \over 2M} + {p^\mu p_\mu  \over 2m} +
V(x),\cr
&\equiv K_{CM} + K_{rel}, \cr}
\eqno(13)$$
where $M= M_1 + M_2$, $m= M_1 M_2 /(M_1 +  M_2)$, and $x = x_1 -x_2$. 
Both $K_{CM}$ and $K_{rel}$
are constants of the motion; the total and relative momenta for the
quantum case may be
represented by partial derivatives with respect to the corresponding 
coordinates.  This problem was solved explicitly for
the classical case by Horwitz and Piron [5], where it was
shown that
there is no precession of the typeP predicted by Sommerfeld [14], who
used the nonrelativistic form $1/r$ for the potential (and obtained a
period for the precession of Mercury that does not fit the data).
\par The  corresponding quantum problem was solved by Cook [15], with
support for the wave functions in the full spacelike region; however, he
obtained a spectrum of the form $1/(n+{1 \over 2})^2$,
with $n$ an integer, that does not agree with the Balmer spectrum for
hydrogen. Zmuidzinas [16], brought to our
attention by P. Winternitz [17]), however,
proved that there {\it is no complete orthogonal set of functions} in the full
spacelike region, and separated the spacelike region into two
submanifolds, in each of which there could be complete orthogonal
sets.  The region for which ${\bf
x}^2 > t^2$, in particular, permits the solution of the differential
equations corresponding to the problem posed by $(2.2)$ by separation
of variables and provides spectra that coincide, up to relativistic corrections, with the
corresponding nonrelativistic probPPPlems with  potentials depending on
$r$ alone. We shall call this sector the RMS (reduced Minkowski
space)[18][19].
\par We may see, moreover, that the RMS carries an important physical
interpretation for the nature of the solutions of the differential
equations by examining the appropriate variables describing the full
spacelike and RMS regions. The full spacelike region is spanned by
$$\eqalign{x^0 &= \rho \sinh\beta,\quad  x^1= \rho \cosh\beta 
\cos\phi \sin\theta \cr 
x^2 &= \rho \cosh \beta \sin \phi \sin \theta,\quad x^3= \rho \cosh \beta
\cos \theta \cr} \eqno(14)$$
over all $\rho$ from $0$ to $\infty$, $\beta$  in $(-\infty,\infty)$,
$\phi$ in $(0, 2\pi)$ and $\theta$ in $(0,\pi)$. Separation of variables in this choice, however, leaves the variable $\beta$ for last; the quantum number (separation constant) obtained in this way has no obvious physical interpretation. Moreover, as found by Cook [15], the resulting spectrum for the Coulomb type potential (proportional to $1/\rho$) does not agree with the Balmer series.    
\par  On the other hand, the set of variables describing the RMS, running over
the same range of parameters [16],
$$\eqalign{x^0 &= \rho \sin \theta \sinh\beta,\quad  x^1= \rho \sin\theta
\cosh\beta \cos\phi 
\cr
x^2 &= \rho \sin \theta \cosh \beta \sin \phi,\quad x^3= 
\rho \cos \theta, \cr} \eqno(15)$$
cover the entire space within the RMS (for
$x_1^2 +x_2^2 >t^2$).
In this coordinatization, the separation constant for $\theta$ (at the last stage), which enters the radial equation and determines the corresponding spectrum, has the interpretation of the angular momentum quantum number $\ell(\ell+1)$.
\par As for $(14)$, for $\beta \rightarrow 0$, these coordinates
become the standard spherical representation of the three dimensional
space (at the ``simultaneity'' point $t=0$, where $\rho$ becomes
$r$). Independently of the form of the
potential $V(\rho)$, one obtains the same radial equation (in $\rho$)
as for the nonrelativistic Schr\"odinger equation (in $r$), and
therefore the same spectra (the two-body {\it mass} squared) for the reduced Hamiltonian. We shall discuss the
relation of these results to the energy spectrum after writing the
solutions. We summarize in the following the basic mathematical steps.
\par Assuming the total wavefunction (for $P\rightarrow P'$, a point
on the continuum of the spectrum of the conserved operator $P$)
$$\Psi_{P'\tau} (X,x)= e^{iP'^\mu X_\mu} \psi_{P'\tau} (x), \eqno(16)$$
the evolution equation
for each value of the total energy momentum of the system is then 
$$ i {\partial \over \partial \tau} \Psi_{P'\tau}(X,x)=
(K_{CM} + K_{rel})\Psi_{P'\tau}(X,x) =
 \bigl[{P'^2
\over 2M} + K_{rel} \bigr]  \Psi_{P'\tau}(X,x). \eqno(17)$$
For the case of discrete eigenvalues $K_a$ of $K_{rel}$.
\par We then have
the eigenvalue equation (cancelling the center of mass wave function
factor and $K_{CM}$ on both sides)
$$ \eqalign{K_{rel} \psi^{(a)}(x)&= K_a \psi^{(a)}(x)\cr
&= (- (1/2m) \partial_\mu\partial^\mu +
V(\rho))\psi^{(a)}(x).\cr} \eqno(18)$$
 Using the $O(3,1)$ Casimir operator, in a way quite
analogous to the the use of the square of the total angular momentum
operator, the Casimir operator of the  rotation group $O(3)$ in the
nonrelativistic case, we may separate the angular and hyperbolic
angular degrees of freedom from the $\rho$ dependence. There are two
Casimir operators defining the representations of $O(3,1)$[20][21][22].
The first
Casimir operator is 
$$\Lambda = {1\over 2} M_{\mu\nu} M^{\mu\nu}; \eqno(19)$$   
the second Casimir operator ${1 \over 2}
\epsilon^{\mu\nu\lambda\sigma}M_{\mu\nu}M _{\lambda\sigma}$ is
identically zero for two particles without spin. Recalling that our
separation into center of mass and relative motion is canonical, and
that 
$$ M^{\mu\nu} = x^\mu p^\nu - x^\nu p^\mu; \eqno(20)$$
using the canonical commutation relations, one finds that
$$ \Lambda = x^2 p^2 + 2ix\cdot p -(x\cdot p)^2. \eqno(21)$$  
Since
$$ x\cdot p\equiv x^\mu p_\mu = -i \rho {\partial \over \partial
\rho}, \eqno(22)$$
so that
$$\Lambda = -\rho^2 \partial^\mu\partial_\mu + 3\rho {\partial \over
\partial \rho} + \rho^2 {\partial^2 \over \partial \rho^2},$$
or 
$$ -\partial_\mu \partial^\mu = -{\partial^2 \over\partial \rho^2}- {3
\over \rho}{\partial \over \partial
\rho} +  {\Lambda \over \rho^2}. \eqno(23)$$
\par Eq. $(18)$ can then be written as
$$ K_a \psi^{(a)}(x) = \bigl\{{1 \over 2m}\bigl[-{\partial^2 \over\partial
\rho^2}- {3\over \rho}{\partial \over \partial \rho} +  
 {\Lambda \over \rho^2}\bigr] + V(\rho)\bigr\}
 \psi^{(a)}(x). \eqno(24))$$
\par  Choosing the RMS variables as
 we have defined them in $(15)$, and with 
$$ L_i = {1 \over 2} \epsilon_{ijk} (x^j p^k - x^k p^j), \eqno(25)$$
corresponding to the definition of the nonrelativistic angular
momentum ${\bf L}$, and
$$ A^i = x^0 p^i -x^i p^0, \eqno(26)$$
corresponding to the boost generator ${\bf A}$, 
$$ \Lambda = {\bf L}^2 - {\bf A}^2. \eqno(27)$$
We then find that 
$$ \Lambda = -{\partial^2 \over \partial \theta^2} - 2 \cot \theta
{\partial \over \partial \theta} + { 1 \over \sin^2 \theta} N^2,
\eqno(28)$$
where 
$$N^2 = L_3^2 - A_1^2 -A_2^2 \eqno(29)$$
is the Casimir operator of the $O(2,1)$ subgroup of $O(3,1)$ leaving
the $z$ axis (and the RMS submanifold) invariant [18].  In terms of the 
 RMS variables that we have defined above,
$$N^2 =  {\partial^2 \over \partial \beta^2} + 2 \tanh \beta {\partial
\over \partial \beta} - {1 \over \cosh^2 \beta} {\partial^2 \over
\partial \phi^2}. \eqno(30)$$
\par We now proceed to separate variables and find the
eigenfunctions. The solution of the general eigenvalue problem
$(24)$ can be written
$$ \psi(x) = R(\rho) \Theta(\theta) B(\beta) \Phi(\phi), \eqno(31)$$
with invariant measure in the $L^2(R^4)$ of the RMS
$$ d\mu = \rho^3 \sin^2 \theta \cosh \beta d\rho d\phi d\beta
d\theta. \eqno(32)$$
\par  To satisfy the $\phi$
derivatives in $(30)$, it is necessary to  take
$$ \Phi_m (\phi) = { 1 \over \sqrt{2\pi}}e^{i[m+{1 \over
2}]\phi},\quad 0\leq \phi <2\pi, \eqno(33)$$
where we have indexed the solutions by the separation constant $m$. For
the case $m$ an integer, this is a double valued function.  To be
compatible with the conditions on the other factors, this is the
necessary choice; one must use, in fact,$\Phi_m (\phi)$ for $m \geq
0$ and $\Phi_m^* (\phi)$ for $m<0$.
\par It has been suggested by M. Bacry [23] that the
occurrence of the half-integer in the phase is associated with the fact
that the RMS is a connected, but not simply connected manifold.  One
can see this by considering the projective form of the restrictions
$$ x^2 + y^2 + z^2 - t^2 >0 \eqno(34)$$
assuring that the events are relatively spacelike, and 
$$ x^2 + y^2 -t^2 >0, \eqno(35)$$
assuring, in addition,  that the relative coordinates lie in the RMS.
Dividing  $(34)$ and $(35)$ by $t^2$, and calling the corresponding
projective variables $X,Y,Z$, we have from $(34)$
$$ X^2 + Y^2 + Z^2 >1, \eqno(36)$$
the exterior of the unit sphere in the projective space, and from $(35)$,
$$ X^2 + Y^2 >1, \eqno(2.30)$$
the exterior of the unit cylinder along the $z$-axis. Identifying the points at infinity of the cylinder,
we see that this corresponds to a torus with the unit sphere
imbedded in the torus at the origin. Such a topological structure is
associated with half integer phase ({\it e.g.}[24]).
 \par We now continue with our discussion of the structure of the
 solutions.
\par The operator $\Lambda$ contains the $O(2,1)$ Casimir $N^2$; with
our solution $(2.23)$, we then have
$$\eqalign{ N^2 B_{mn} (\beta) &= 
 \bigl[  {\partial^2 \over \partial \beta^2} + 
2 \tanh \beta {\partial\over \partial \beta} 
+{(m+ {1\over2})^2 \over \cosh^2 \beta}\bigr]B_{mn}(\beta)\cr
&\equiv (n^2 - {1 \over 4})B_{mn}(\beta),\cr}  \eqno(38)$$
 where $n^2$ is the separation constant for the variable $\beta$.  The 
 term $(m+ {1 \over 2})^2$ must be replaced by $(m-{1\over 2})^2 = (|m|
 + {1 \over 2})^2$ for $m<0$.  We study only the case $m\geq 0$ in
 what follows. The remaining equation for $\Lambda$ is then
$$ \Lambda \Theta(\theta) = \bigl[  -{\partial^2 \over \partial
\theta^2} - 2 \cot \theta
{\partial \over \partial \theta} + { 1 \over \sin^2 \theta}\bigl( n^2
- {1 \over 4}\bigr)\bigr]\Theta(\theta). \eqno(39)$$ 
\par For the treatment of Eq $(38)$, it is convenient to make the
substitution
$$ \zeta = \tanh\beta, \eqno(40)$$
so that $-1 \leq \zeta \leq 1$.  One then finds that for 
$$B_{mn}(\beta) = (1-\zeta^2)^{1/4} {\hat B}_{mn} (\zeta),
\eqno(41)$$
 $(38)$ becomes 
$$ \eqalign{(1-\zeta^2) &{\partial^2 {\hat B}_{mn} (\zeta) \over \partial
\zeta^2} - 2\zeta {\partial {\hat B}_{mn} (\zeta) \over \partial
\zeta}\cr
&+ \bigl[ m(m+1) - {n^2 \over 1-\zeta^2 }\bigr] {\hat B}_{mn}(\zeta)
=0. \cr}\eqno(42)$$ 
The solutions are the associated Legendre functions of the first and
second kind \break\hfill (Gel'fand[21]; see also Merzbacher[25]),
 $P_m^n (\zeta)$ and $Q_m^n(\zeta)$. The normalization
condition on these solutions, with the measure $(42)$ is
$$ \int \cosh \beta |B(\beta)|^2 <\infty, $$
or, in terms of the variable $\zeta$
$$ \int_{-1}^1 (1- \zeta^2)^{-1} |{\hat B}(\zeta)|^2 d\zeta <
\infty. \eqno(43)$$ 
The second kind Legendre functions do not satisfy this condition. For
the condition on the $P_m^n (\zeta)$, it is simplest to write
the known result [26]
$$\int_{-1}^1 (1 - \zeta^2)^{-1} |P_{\mu + \nu}^{-\nu}(\zeta)|^2
d\zeta = { 1 \over \nu} {\Gamma (1 + \mu) \over \Gamma(1 + \mu +2\nu)}
\eqno(44)$$
The normalized solutions (it is sufficient to consider $n\geq 0$) may
be written as
$$ {\hat B} _{mn}(\zeta) = \sqrt{n} \sqrt{[\Gamma(1+m+n) /\Gamma(1+ m
-n)]}\times P_m^{-n}(\zeta), \eqno(45)$$
where $m \geq n$.
\par The
case $n=0$ must be treated with special care; it requires a
regularization. For $n=0$, the associated Legendre functions become 
the Legendre polynomials $P_m(\zeta)$.   In terms of the integration
on $\beta$, the factor $\cosh \beta = (1-\zeta^2)^{-1/2}$ in the
measure is cancelled by the square of the factor $(1-\zeta^2)^{1/4}$
in the norm, so that the integration appears as
$$ \int_{-\infty}^\infty |{\hat B}_m(\zeta)|^2 d\beta.$$
  The Legendre
polynomials do not vanish at $\zeta=\pm 1$, so if ${\hat B}_m$ and $P_m$
are related by a finite coefficient, the integral would diverge. When
$n$ goes to zero, associated with the ground state,
the wave function spreads along the hyperbola labelled by $\rho$,
going asymptotically to the light plane; the probability  density
with respect to intervals of $\beta$ becomes constant for large
$\beta$.  The (regularized) expectation values reproduce the 
distribution of the Schr\"odinger bound states, although the 
spacetime wave function
approaches that of a generalized eigenfunction.
\par To carry out the regularization, we take the
limit as $n$ goes continuously to zero after computation of scalar
products.  Thus, we assume the form
 $$ {\hat B}_m(\zeta) = \sqrt{\epsilon} (1-\zeta^2)^{\epsilon/2} P_m
 (\zeta), \eqno(46)$$
with $\epsilon \rightarrow 0$ after computation of scalar
 products.  This formula is essentially a residue of the Rodrigues formula
$$P_m^{-n}(\zeta) = (-1)^n (1- \zeta^2)^{n/2} {d^n \over d\zeta^n}
P_m(\zeta) \eqno(47)$$
for $n \rightarrow 0$.  
\par  The operator for the differential equation $(24)$ for
the eigenvalue of the reduced motion is invariant under the action of
the Lorentz group. It follows from acting on the equation with the
unitary representation of the Lorentz group that the eigenfunctions
must be representations of that group [24] for each value
of the eigenvalue.  However, as one can easily see, the solutions that
we found are, in fact, irreducible representations of $O(2,1)$, not,
{\it a priori}, representations of the Lorentz group $O(3,1)$. We discuss below how to construct such a representation.
\par We have required that the wave functions be eigenfunctions of the
Casimir operator $(29)$ of the $O(2,1)$ subgroup. For the generators
of $O(2,1)$, we note that
$$\eqalign{ H_\pm \equiv A_1 \pm iA_2 &= e^{\pm i\phi}\bigl( -i {\partial \over
\partial \beta} \pm \tanh\beta {\partial \over \partial \phi}
\bigr),\cr
L_3 &= -i {\partial \over \partial \phi},\cr
A_3 &= -i\bigl(\cot \theta \cosh\beta {\partial \over \partial \beta} -
\sinh\beta {\partial \over \partial \theta}\bigr)\cr
L_\pm &= L_1 \pm iL_2\cr
 &= e^{\pm i \phi}\bigl( \pm \cosh\beta {\partial
\over \partial \theta } - \sinh \beta  \cot \theta {\partial \over
\partial \beta} \cr
&+ i {\cot \theta \over \cosh \beta} {\partial \over \partial \phi}
\bigr).\cr} \eqno(48)$$
It then follows that $H_{\pm}$ are raising and lowering operators for $m$
on the functions 
$$ \eqalign{\xi_{n+k}^{-n} (\zeta, \phi) &\equiv B_{n+k,n} (\beta)
\Phi_{n+k}(\phi)\cr 
&= (1-\zeta^2)^{1/4} {\hat B}_{n+k,n}(\zeta) \Phi_{n+l} (\phi), \cr }
\eqno(49)$$ 
where it is convenient to replace $m$ by $n+k$. With the relation
$$ [L_3, H_\pm] = \pm H_\pm \eqno(50)$$
one can show [19] that
$$ H_+ \chi_{n+k}^{-n}(\zeta, \phi) = i\sqrt{(k+1)(2n
+k+1)}\chi_{n+k+1}^{-n} (\zeta, \phi) \eqno(51)$$
and that
$$  H_- \chi_{n+k+1}^{-n}(\zeta, \phi) = -i\sqrt{(k+1)(2n
+k+1)}\chi_{n+k}^{-n} (\zeta, \phi).\eqno(52)$$
The complex conjugate of $\chi_{n+k}^{-n}$ transforms in a similar
way, resulting in a second (inequivalent) representation of
$O(2,1)$ with the same value of the $O(2,1)$ Casimir operator (these
states correspond to replacement of $m+{1\over 2}$ by $m-{1\over 2}$
for $m <0$, and are the result of charge conjugation.  Since the
operators $A_1, A_2$ and $L_3$ are Hermitian, complex conjugation is
equivalent to the transpose.  Replacing these operators by their
negative transpose (defined by $C$), leaves the commutation
relations invariant. Thus the action on the complex conjugate states
involves
$$\eqalign{H_-^C &= -H_+^* = H_-, \quad H_+^C = -H_-^* = H_+, \cr
L_3^C &= -L_3^* = L_3;\cr} \eqno(53)$$
 These are precisely the operators under which the complex conjugate
 states transform, and this operation therefore corresponds to charge conjugation.
 \par The wave functions we have
obtained are irreducible representations of $O(2,1)$, determined by the differential equations with solutions retricted to support in a particular choice of orientation of the RMS. To construct
representations of $O(3,1)$, let us consider first the  well established method
which is effective in constructing representations of $O(3,1)$ from
representations of $O(3)$, a group that we would have found if we were
working with solutions in the timelike region [21],
called the {\it ladder representation}.  It follows from the Lie
algebra of $O(3,1)$ that the $O(3)$ subgroup Casimir operators
$\ell(\ell+1)$ are stepped by $\ell \rightarrow \ell \pm 1$
 under the action of the boost from $O(3,1)$.  The whole
set of representations of $O(3)$, from $\ell=0$ to $\infty$ form a
representation of $O(3,1)$. Each of the representations of $O(3)$
entering this tower are trivially normalizable, since they are of
dimension $(2 \ell +1)$.  However, attempting to apply this method to
the representations of $O(2,1)$ fails because
the application of the Lie algebra to this set connects the lowest 
state of the tower with the ground state which, as we have shown,
requires regularization. The action of the algebra does not provide such
a regularization, and therefore the method is inapplicable.
\par We therefore turn to the method of {\it induced representations} [27]. We may apply this method to contructing the representations of
$O(3,1)$ based on an induced representation with the $O(2,1)$ ``little
group'', based on a spacelike vector corresponding to the choice of
the $z$ axis.   We shall discuss his method in detail below.
\par We first record the solutions of the equation $(18)$.
\par Defining 
$$ \xi = \cos \theta \eqno(54)$$
and the functions
$$ {\hat \Theta}(\theta) = (1-\xi^2)^{1/4} \Theta(\theta), \eqno(55)$$
Eq. $(39)$ becomes
$${d\over d\xi} \bigl( (1-\xi^2) {d \over d\xi} {\hat \Theta}(\theta)
\bigr) + \bigl( \ell(\ell +1) - {n^2 \over 1-\xi^2} \bigr) {\hat
\Theta}(\theta)=0 ,\eqno(56)$$  
where we have defined 
$$ \Lambda = \ell(\ell+1) - {1 \over 4} . \eqno(57)$$
The solutions are proportional to the associated Legendre
functions of the first or second kind, $P_\ell^n(\xi)$ or
$Q_\ell^n(\xi)$.  For $n \neq 0$, the second kind functions are not
normalizable. We therefore reject these. 
\par The normalizable irreducible representations of $O(2,1)$ are single or
double valued, and hence $m$ must be integer or half integer. As we
have seen, $k$ is integer valued, and therefore $n$ must be integer or
half integer also. Normalizability conditions on the associated
Legendre functions then require that $\ell$ be respectively, positive
half-integer or integer.  The lowest mass state, as we shall see from
the spectral results, corresponds to $\ell =0$, and hence we shall
consider only integer values of $\ell$.Therefore, $n$ and $m$ must be integer. 
\par We now turn to the solution of the radial equations, containing
the spectral content of the theory. With the evaluation of $\Lambda$
in $(57)$, we may write the radial equation as
$$ \eqalign{\bigl[ { 1 \over 2m} \bigl( &-{\partial^2 \over \partial \rho^2} -
{3 \over \rho} {\partial \over \partial \rho} + {\ell (\ell +1)  - {3
\over 4} \over \rho^2}\bigr) + V(\rho) \bigr]R^{(a)}(\rho)\cr & = K_a 
 R^{(a)}(\rho).\cr} \eqno(58)$$ 
\par If we put
$$ R^{(a)}(\rho) = { 1 \over \sqrt{\rho}} {\hat R} ^{(a)}(\rho),
\eqno(59)$$
Eq. $(58)$ becomes precisely the nonrelativistic Schr\"odinger
equation for ${\hat R}^{(a)}$ in the variable $\rho$, with potential
$V(\rho)$ (the measure for these functions is, from $(32)$, just
$\rho^2 d \rho$, as for the nonrelativistic theory)

$$\eqalign{{d^2 {\hat R}^{(a)} (\rho) \over d\rho^2} &+ {2 \over \rho}{ d{\hat
R}^{(a)}(\rho)\over d\rho} \cr &- {\ell(\ell +1) \over \rho^2} {\hat
R}^{(a)}(\rho) \cr &+ 2m (K_a - V(\rho)){\hat R}^{(a)}(\rho) = 0.\cr}
 \eqno(60)$$
 \bigskip
 \noindent{\bf 3. The spectrum}
 \bigskip
\par The lowest eigenvalue $K_a$, as for the energy in the nonrelativistic
Schr\"odinger equation, corresponds to the $\ell = 0$ state of the
sequence $\ell = 0,1,2,3,...$, and therefore the quantum number
$\ell$ plays a role analogous to the orbital angular momentum.  This
energy is of a lower value than achievable with wave functions with
support in the full spacelike region [15] and the relaxation of
the system to wave functions with support in the RMS may be thought of,
in this sense, as a spontaneous symmetry breaking (we thank A. Ashtekar
for his remark on this point [28]).
\par The value of the full generator $K$ is then determined by these
eigenvalues and the value of the center of mass total mass squared
operator, {\it i.e.}, 
$$ K = {P^\mu P_\mu \over 2M} + K_a. \eqno(61)$$
The first term corresponds to the total effective rest mass of the
system. In particular, the invariant mass squared
of the system is given by (sometimes called the Mandelstam variable
$s$[29])
$$ s_a \equiv -P_a^2 = 2M(K_a -K). \eqno(62)$$ 
This total center of mass momentum is observed in the laboratory in
scattering and decay processes, where it is defined as the sum of the
outgoing momenta squared. In the case of two particles, it would be
given by $-(p_1^\mu + p_2^\mu)(p_{1\mu} + p_{2\mu})$, as we have
defined it in $(62)$. This quantity is given in terms of total
energy and momentum by
$$ s_a = E_T^2 -{\bf P}_T^2, \eqno(63)$$
and in the center of momentum frame, for ${\bf P} =0$,  is just $E_T^2$.
\par In order to extract information about the
{\it energy} spectrum,we must therefore make some assumption on the
value of the  conserved quantity $K$. In the case of a potential that
vanishes for large $\rho$, we may consider the two particles to be
asymptotically free, so the effective Hamiltonian in this asymptotic region
$$ K \cong {{p_1}^\mu {p_1}_\mu \over 2M_1} +  {{p_2}^\mu {p_2}_\mu \over
2M_2}.\eqno(64)$$
Further, assuming that the two particles at very large distances, in
accordance with our experience, undergo a relaxation to their mass
shells, so that $p_i^2 \cong -M_i^2$.  In this case, $K$ would be assigned the value
$$ K\cong -{M_1 \over 2} - {M_2 \over 2} = - {M\over 2}. \eqno(65)$$
The two particles in this asympotic state would, for the bound state
problem, be at the ionization point.  If these
assumptions are approximately valid, we find for the total energy,
which we now label $E_a$,
$$ E_a/c \cong \sqrt{M^2c^2 + 2MK_a}, \eqno(66)$$
where we have restored the factors $c$.
\par In the case of excitations small compared to the total mass of the
system, we may factor out $Mc$ and represent the result in a power
series expansion
$$ E_a \cong Mc^2 + K_a  - {1 \over 2} {K_a \over Mc^2} + \dots,
\eqno(67)$$
so that the energy spectrum is just the set $\{K_a\}$ up to relativistic
corrections. Thus, the spectrum for the $ 1/\rho$ potential is just
that of the nonrelativistic hydrogen problem up to relativistic
corrections, of order $ 1/c^2$.
 \par If the spectral set  $\{K_a\}$ includes large negative values,
 the result $(66)$ could become imaginary, indicating the possible
 onset of instability.  However, the asymptotic condition imposed on
 the evaluation of $K$ must be re-examined in this case. If the
 potential grows very rapidly as $\rho \rightarrow 0$, then at large
 spacelike distances, where the hyperbolic surfaces  $\rho = const$
 approach the lightcone, the Euclidean measure $d^4x$ (thought of, in
 this context,  as
 small but finite) on the $R^4$ of
 spacetime starts to cover very singular values and the
 expectation values of the Hamiltonian at large spacelike distances
 may not permit the contribution of the potential to become
 negligible; it may have an effectively very long range. This effect
 can occur in the
 transverse direction to the $z$ axis along the tangent to the light
 cone; the hyperbolas cannot reach the light cone in the $z$
 direction, which may play an
 important role in the modelling the behavior of the transverse
 scattering amplitudes
 in high energy scattering studied, for example, by Hagedorn[30].
 \bigskip
\noindent{\bf 4. Some examples}
\bigskip
\par In this section we give the examples of the Coulomb
potential and the oscillator.
\par For the analog of the Coulomb potential, we take
$$ V(\rho) = -{Ze^2 \over \rho}. \eqno(68)$$
As we have remarked above,for $c \rightarrow \infty$, this potential reduces to $1/r$, the usual Coulomb, and therefore the spectrum must reduce to the usual Balmer series in this limit. 
\par In this case the spectrum, according to the solutions above, is given
by
$$K_a = -{Z^2 me^4 \over 2\hbar^2 (\ell +1 +n_a)^2}, \eqno(69)$$
where $n_a = 0,1,2,3....$.  The wave functions ${\hat R}(\rho)^a $ are
the usual hydrogen functions
 $$ {\hat R}_{n_a \ell} (\rho) =\sqrt{ {Z n_a! \over (n_a + \ell +1)^2 (n_a
 + 2\ell +1)}} e^{-x/2} x^{\ell +1} L_{n_a}^{2\ell +1} (x),
 \eqno(70)$$
where $L_{n_a}^{2\ell +1}$ are the Laguerre polynomials, and the
 variable $x$ is defined by
$$ x = {(2Z \rho/ a_0) \over (n_a + \ell +1)}, \eqno(71)$$
and $a_0 = \hbar^2 / me^2)$.  The {\it size} of the bound state, which
is related to the atomic form factor, is measured according to the
variable $\rho$ [31].  For the lowest level  (using the 
regularized functions) $n_a = \ell=0$, 
$$<\rho>_{n_a = \ell = 0} = {3\over 2} a_0. \eqno(72)$$
The total mass spectrum, given by $(62)$, is then
$$ s_{n_a, \ell} \cong M^2c^2 - {mM Z^2 e^4 \over \hbar^2 (n_a +\ell
+1)^2}. \eqno(73)$$
For the case that the nonrelativistic spectrum has value small
compared to the sum of the particle rest masses, we may  use the
approximate relation $(66)$ to obtain
$$\eqalign{ E_{a,\ell}\cong Mc^2 &- {Z^2 m e^4 \over 2\hbar^2 (n_a
+\ell +1)^2}\cr &- {1 \over 8} {Z^4 m^2 e^8 \over Mc^2 \hbar^4 (n_a
+\ell +1)^4} +\dots. \cr} \eqno(74)$$
The lowest order relativistic correction to the rest energy of the two
body system with Coulomb like potential is then 
$$ {\Delta (E_{a, \ell} -Mc^2) \over E_{a, \ell} -Mc^2} = {Z\alpha^2
\over 4} \bigl( {m \over M}\bigr)  { 1 \over (n_a + \ell +
1)^2}. \eqno(75)$$
  For positronium,   $\Delta (E -Mc^2)\sim
2\times 10^{-5}$ eV it is about one part in $10^5$, about $2\%$ of the
positronium hyperfine splitting  of $8.4 \times 10^{-4}$ eV [32]. 
We see quantitatively that the relativistic
theory gives results that are consistent with the known data on these
experimentally well studied bound state systems.
\par For the four dimensional  oscillator, with $V(\rho)= {1
\over2}m\omega^2 \rho^2$, Eq.$(60)$ takes the form
 $$\eqalign{{d^2 {\hat R}^{(a)} (\rho) \over d\rho^2} &+ {2 \over \rho}{ d{\hat
R}^{(a)}(\rho)\over d\rho} \cr &- {\ell(\ell +1) \over \rho^2} {\hat
R}^{(a)}(\rho) \cr &+ 2m \bigl(K_a - {m^2 \omega^2 \over \hbar^2} \rho^2 -
{\ell(\ell +1)\over \rho^2}\bigr){\hat R}^{(a)}(\rho) = 0.\cr}
  \eqno(76)$$
With the transformation
$$ {\hat R}^{(a)} (\rho) = x^{\ell/2} e^{-x/2} w^{(a)} (x), \eqno(77)$$
for 
$$ x = { m\omega \over \hbar} \rho^2, \eqno(78)$$
we obtain the equation
$$ \eqalign{x {d^2 w^{(a)}\over dx^2} &+ \bigl( \ell + {3 \over 2} -x \bigr)
{dw^{(a)} \over dx}\cr  &+ {1 \over 2} \bigl(\ell + {3\over 2} - {K_a
\over\hbar \omega} \bigr) w^{(a)}=0\cr} \eqno(79)$$
Normalizable solutions, the Laguerre polynomials $L_{n_a}^{\ell +
1/2}(x)$,  exist [18] when the coefficient of  
$w^{(a)} (x)$ is a negative integer, so that the eignvalues are 
$$ K_a = \hbar \omega(\ell + 2n_a + {3\over 2}), \eqno(80)$$
where $n_a = 0,1,2,3,\dots$  The total mass spectrum is given by
$(62)$ as
$$ s_{n_a, \ell} = -2MK + 2M\hbar \omega(\ell + 2n_a + {3\over
2}),\eqno(81) $$
Note that the ``zero point'' term is ${3 \over 2}$, indicating that in
the RMS, in the covariant equations there are effectively three intrinsic
degrees of freedom, as for the nonrelativistic oscillator. 
\par The choice
of $K$ is arbitrary here, since there is no ionization point for the
oscillator, and no {\it a priori} way of assigning it a value;
setting $K = -{Mc^2 \over 2}$ as for the Coulomb problem
(a choice that may be justified by setting the spring constant equal
to zero and adiabatically increasing it to its final value), one
obtains, for  small excitations relative to the particle masses,
 $$ \eqalign{E_a \cong Mc^2 &+ \hbar \omega \bigl(\ell + 2n_a + 
{ 3\over 2}\bigr)\cr &- {1 \over 2} {\hbar^2 \omega^2(\ell +2n_a + {3
\over 2} )^2 \over Mc^2} +\dots \cr} \eqno(82)$$
Feynman, Kislinger and Ravndal [33], Kim and Noz
[34] 
and Leutwyler and Stern [35]
have studied the relativistic oscillator and obtained a positive
spectrum by imposing a subsidiary condition
suppressing timelike excitations, which lead, in the formalism of
annihilation-creation operators to generate the spectrum, to negative
norm states (``ghosts''). There are no ghost states in the covariant
treatment we discuss here, and no extra constraints invoked in finding the
spectrum. The solutions are given in terms of Laguerre poynomials, but
unlike the case of the standard treatment of the $4D$ oscillator, in
which $x^\mu \pm i p^\mu$ are considered annihilation-creation
operators, the spectrum generating algebra (for example, Dothan [36]) for the covariant SHP
oscillator has been elusive [37].
\bigskip
\noindent{\bf 5 The induced representation}
\bigskip
\par We have remarked that the solutions of the invariant two body
problem results in solutions that are irreducible representations of
$O(2,1)$, in fact, the complex representations of its covering group
$SU(1,1)$, and pointed out that the ladder representations generated
by the action of the Lorentz group on these states cannot be used to
obtain representations of the full Lorentz group $O(3,1)$ or its
covering $SL(2,C)$. Since the differential equations defining the
physical states  are covariant under the action of $O(3,1)$, the
solutions must be representations of $O(3,1)$.  To solve this problem,
one observes [1] that the $O(2,1)$ solutions are constructed in the RMS
which is referred to the spacelike $z$ axis.  Under a Lorentz boost,
the entire RMS turns, leaving the light cone invariant. After this transformation the new RMS is constructed on the
basis of a new spacelike direction which we call here $m^\mu$. However, the differential equations
remain identically the same since the
operator form of these equations is invariant undr Lorentz transformations.. The change of coordinates to RMS variables has the same form as well, and therefore the set of solutions of these equations have the same structure.  These functions are now
related to the new $z$ axis. Under the action of the full 
Lorentz group the wave functions undergo a transformation involving a linear
combination of the set of eigenfunctions found in the previous
section; this action does not change the value of the $SU(1,1)$ (or
$O(2,1)$) Casimir operator; together with the change in direction of
the vector $m^\mu$, they provide an induced representation of $SL(2,C)$ (or
$O(3,1)$ with little group $SU(1,1)$ in the same way that relativistic spin is a
representation of $SL(2,C)$ with $SU(2)$ little group [27].
 \par Let us define the coordinates $\{y_\mu\}$, isomorphic to the set
$\{x_\mu\}$, defined in an accompanying frame for the RMS($m_\mu)$),
with $y_3$ along the axis $m_\mu$. Along with infinitesimal operators
of the $O(2,1)$ generating changes within the  RMS($m_\mu)$), there
are generators on $O(3,1)$ which change the direction of $m_\mu$; as
for the induced representations for systems with spin [27], the Lorentz group
contains these two actions, and therefore both Casimir operators are
essential to defining the representations,{\it i.e.}, both
$$ c_1 \equiv {\bf L}(m)^2 - {\bf A }(m)^2 \eqno(83)$$ 
and
$$ c_2 \equiv {\bf L}(m)\cdot {\bf A }(m), \eqno(84)$$
which is not identically zero, and commutes with $c_1$.
\par In the following, we construct functions on the {\it orbit} of
the $SU(1,1)$ little group representing the full Lorentz group; along
with the designation of the point on the orbit, labelled by $m_\mu$,
these functions constitute a description of the physical state of the
system.
\par It is a quite general result that the induced representation of a
noncompact group contains all of the irreducible representations.  We
decompose the functions along the orbit into basis sets
corresponding to eigenfunctions for the $O(3)$ subgroup Casimir operator
${\bf L}(m)^2\rightarrow L(L+1)$  and $L_1 \rightarrow q$ that take on
values that persist along the orbit; these solutions correspond to the
{\it principal series } of Gel'fand [21]. These quantum
numbers 
for the induced representation do not correspond directly to the observed
angular momenta of the system. The values that correspond to spectra
and wavefunctions with nonrelativistic limit coinciding with those of
the nonrelativistic problem problem, are those with $L$ {\it
half-integer} for the lowest Gel'fand $L$ level. The
Gel'fand classification, the two Casimir operators take on the values
$c_1 = L_0^2 + L_1^2 -1, \quad c_2 = -iL_0 L_1 $, where $L_1$ is pure
imaginary and , in general, $L_0$ is integer or half-integer. In the
nonrelativistic limit, the action of the group on the relative
coordinates becomes deformed in such a way that the $O(3,1)$ goes
into the nonrelativistic $O(3)$, and the $O(2,1)$ into the $O(2)$
subgroup in the initial configuration of the RMS based on the $z$ axis.  
\par The representations that we shall obtain, in the principal series
of Gel'fand [21], are unitary in a Hilbert space with
scalar product that is defined by an integration invariant under the full
$SL(2,C)$, including an integration over the the measure  space of
$SU(1,1)$, carried out in the scalar product in $L^2( R^4 \subseteq
{\rm RMS}(m_\mu))$, for each $m_\mu$ (corresponding to the orientation
of the new $z$ axis, and an integration over the measure of the coset
space $SL(2,C)/ SU(1,1)$; the complete measure is $d^4 y d^4 m \delta
(m^2 -1) $, {\it i.e.}, a probability measure on $R^7$, where $y_\mu
\in {\rm RMS}(m_\mu)$.  The coordinate description of the quantum
state therefore corresponds to an ensemble of (relatively defined)
events lying in a set of RMS($m_\mu$)'s over all possible spacelike
$\{m_\mu\}$. 
\par A coordinate system oriented with its $z$ axis along the
direction $m_\mu$, as referred to above, can be constructed by means
of a coordinate transformation of Lorentz type (here $m$ represents
the spacelike orientation of the transformed RMS, not to be confused
with a magnetic quantum number),
$$ y_\mu = {L(m)_\mu}^\nu x_\nu.   \eqno(85)$$
\par  For example, if we take a vector $x_\mu$ parallel to $m_\mu$, with
 $x_\mu = \lambda m_\mu$, then the corresponding $y_\mu$ is $\lambda
 m^0_\mu$, with $m^0_\mu$ in the direction of the initial orientation
 of the orbit, say, the $z$ axis. This definition may be replaced by
 another by right multiplication of an element of the stability group
 of $m\mu$ and left multiplication by an element of the stability
 group of $m^0_\mu$, constituting an isomorphism in the RMS.
\par The variables $y_\mu$ may be parametrized by the same
 trigonometric and hyperbolic functions as in $(15)$ since they span
 the RMS, and provide a complete characterization of the configuration
 space in the RMS($m_\mu$) that is universal in the sense that it is
 the same in every Lorentz frame.  It is convenient to
 define the functions 
$$\psi_m(y) = \phi_m(L^T(m)y) = \phi_m(x) \eqno(86)$$
\par We can then define the map of the Hilbert
spaces associate with each $m_\mu$ in the foliation ${\cal H}_m
\rightarrow {\cal H}_{\Lambda m}$ such that the state vectors are
related by the norm preserving transformation         
  $$ \Psi^\Lambda_{\Lambda m} = U(\Lambda) \Psi_m. \eqno(87)$$
In the new Lorentz frame (with $y = L(\Lambda m)x$),
$$ \eqalign{\phi^\Lambda_{\Lambda m} (x)  &= {}_{\Lambda
m}<x|\Psi^\Lambda_{\Lambda m}>\cr
&= {}_{\Lambda m}< x| U(\Lambda) \Psi_m > = \phi^\Lambda_{\Lambda
m}(L^T (\Lambda m)y) \cr
&= \psi^\Lambda _{\Lambda m}(y).\cr} \eqno(88)$$
\par If $\phi_m(x)$ is scalar under Lorentz transformation, so that
(we assume no additional phase) 
 $$ \phi^\Lambda_{\Lambda m} (\Lambda x) = \phi_m(x), \eqno(89)$$
it follows from $(88)$ that 
$$ U(\Lambda) |x>_m = |\Lambda x>_{\Lambda m}. \eqno(90)$$
The wave function $\phi^\Lambda_{\Lambda m} (x)$ describes a system in
a Lorentz frame in motion with respect to the frame in which the
state is described by $\phi_m(x)$, and for which the support is in the
RMS($(\Lambda m )_\mu$).  The value of this function at $x$ in the new
frame is determined by its value at $\Lambda^{-1} x$ in the original
frame; moreover, the subensemble associated with values of $m_\mu$
over the orbit in the new frame is determined by the subensemble
associated with the values of $(\Lambda^{-1}m)_\mu$ in the old frame.
We define the description of the state of the system in the new frame
in terms of the set (over $\{m_\mu\}$) of transformed wave functions
$$ \eqalign{\psi^\Lambda_m(y)&\equiv  \phi_{\Lambda^{-1}m}(\Lambda^{-1}x)\cr
 &= \psi^\Lambda_m(D^{-1}(\Lambda, m)y)\cr} \eqno(91)$$
where we have used $(88)$ (the transformed function has support
 oriented with $m_\mu$) and defined the (pseudo) orthogonal matrix (we
 define a ``matrix'' $A$ as $\{{A_\mu}^\nu \}$) 
$$D(\Lambda,m) = L(m)\Lambda L^T (\Lambda^{-1} m).\eqno(92)$$
  The transformation
$D^{-1}(\Lambda,m)$ stabilizes $m^0_\mu$, and is therefore in the
$O(2,1)$ subgroup that leaves the RMS of the original system
invariant. Eq. $(91)$ defines an induced representation of
$SL(2,C)$, the double covering of $O(3,1)$. 
\par Classification of the orbits of the induced representation are
determined by the Casimir operators of $SL(2,C)$, defined as
differential operators on the functions $\psi_m(y)$ of $(86)$, {\it
i.e.}, the operators defined in $(83)$ and $(84)$.  To define
these variables as differential operators on the space $\{y\}$, we
study the infinitesimal Lorentz transformations
$$\Lambda \cong 1 + \lambda, \eqno(93)$$ 
for which 
$$ {\psi^{1+\lambda}}_m (y) = \psi_{m-\lambda m} (D^{-1} (1 +
\lambda,n)y),\eqno(94)$$
and $\lambda$ is an infinitesimal Lorentz transformation
(antisymmetric).  To first order,the little group transformation is
$$D^{-1} (1+\lambda,n) \cong 1 - (d_m(\lambda)L(m))L^T(m) - L(m)
\lambda L^T(m), \eqno(95)$$

where $d_m$ is a derivative with respect to $m_\mu$ holding $y_\mu$
fixed,
$$ d_m(\lambda) = {\lambda_\mu}^\nu m_\nu {\partial \over \partial
n_\mu}. \eqno(96)$$
From the property $L(m)L^T(m) =1$, it follows that
$$(d_m(\lambda)L(m))L^T(m)= -L(m)(d_m(\lambda)L^T(m)),\eqno(97)$$
so that $(95)$ can be written as
$$\eqalign{D^{-1} (1+\lambda,n) &\cong 1 + L(m) (d_n(\lambda)L^T(m)
-\lambda L^T(m))\cr 
&\equiv 1 -G_m(\lambda). \cr} \eqno(98)$$
For the transformation of $\psi_m$ we then obtain
$$ {\psi^{1+\lambda}}_m (y) \cong \psi_m(y) - d_m(\lambda +
g_m(\lambda))\psi_m(y), \eqno(99)$$
where
$$g_m (\lambda) = {G_m(\lambda)_\mu}^\nu y_\nu {\partial \over \partial
y_\mu}.\eqno(100)$$
Eq. $(99)$ displays explicitly the effect of the transformation
along the orbit and the transformation within the little group. 
\par The algebra of these generators of the Lorentz
group are investigated in [1]; the closure of this algebra
follows from the remarkable property of compensation for the
derivatives of the little group generators along the orbit (behaving
in a way similar to a covariant derivative in differential geometry).  The general
structure we have exhibited here is a type of fiber bundle, sometimes called a 
{\it Hilbert bundle}, consisting of a set of Hilbert spaces on the
base space of
the orbit; in this case, the fibers, corresponding to these Hilbert
spaces, transform under the little group $O(2,1)$.
\par  There are functions on the orbit with definite values of the two
Casimir operators, as well as ${\bf L}(m)^2$ and $L_1(m)$; one finds
the  Gel'fand Naimark canonical represention with decomposition over
the $SU(2)$ subgroup of $SL(2,C)$, enabling an identification of the
angular momentum content of the representations [17].  
 With a consistency relation
between the Casimir operators (for the solution of the finite set of
equations involving functions on the hyperbolic parameters of the
spacelike four vector $m_\mu$), we find
that we are dealing with the principal series of Gel'fand [20][21].
\bigskip
\noindent{\bf Conclusions}
\bigskip
\par We have reviewed and discussed the symmetry of the two body central potential problem in the relativistically covariant framework of the SHP theory. The solutions of the Stuekelberg-Schr\"odinger equation with support in the full spacelike region of the Minkowski space provide a spectrum that does not agree with the solutions of the nonrelativistic Schr\"odinger equation. Guided by the work of Zmuidzinas [15] we used variables in the Minkowski configuration space that span a spacelke $O(2,1)$ invariant subspace of the full Minkowski space for the relative coordinates. In this subspace the spectrum agrees, up to relativistic corrections, with the non-relativistic Schr\"odinger spectrum.
\par In this subspace, which we call the RMS (reduced Minkowski space), the eigenfunctions of the stationary Stueckelberg-Schr\"odiger equation form representations of the orthogonal group $O(2,1)$. However, the Hamiltonian operator is $O(3,1)$ invariant, which implies that the solutions must be representations of $O(3,1)$. Extending the $O(2,1)$ representations by the method of stepping to get a ladder representation leads to a non-normalizable state, and we therefore turned to an {\it induced representation}[19]. This represention was constructed following Wigner's method [27] for dealing with spin in a relativisic framework, but with the non-compact $O(2,1)$ little group instead of the $O(3)$ little group used by Wigner to describe spin. One might think of the reduced symmetry $O(2,1)$, as suggested by Ashtekar[28] as a spontaneous symmetry breaking (the ground state has lower energy than for the solutions in the full spacelike region). This consruction leads to eigenfunctions for the two body problem that have the intrinsic spinorial property of being double valued, perhaps a reflection of the topological properties of the $O(2,1)$ invariant submanifold [23].
\bigskip
\noindent{\bf Acknowledgements}
\par We are greatful to Shmuel Nussinov, Yakir Aharonov, Yeshayahu Lavie [Z''L], Martin Land, Fritz Rohrlich, Abbay Ashtekhar and Louis Michel [Z''L] for helpful discussions.
\bigskip
\centerline{\it References}
\bigskip
\frenchspacing
{\obeylines
[1] Lawrence P. Horwitz, {\it Relativistic Quantum Mechanics}, Springer, Dordrecht (2015).

[2] S. Sklarz and L.P. Horwitz, Relativistic mechanics of continuous media, Found. Phys. {\bf 31}, 909 (2001).

[3] L.P. Horwitz, S. Shashoua and W.C.Schieve, A manifestly covariant relativistic Boltzmann equation for the evolution of a system of events, Physica A {\bf 161}, 300 (1989).

[4] Lawrence P. Horwitz and Rafael I. Arshansky, {\it Relativistic Many Body Theory and Statistical Mechanics}, Morgan and Claypool, IOP Concise Physics, San Rafael CA (2018).

[5]  L.P. Horwitz and C. Piron, Relativistic dynamics, Helv. Phys. Acta {\bf 66}, 316
(1973)

[6] J.R. Fanchi and R.E. Collins, Quantum mechanics of relativistic spinless particles, Foundations of Physics {\bf 8}, 851 (1978).}

{\obeylines
[7] J.R. Fanchi, {\it Parametrized Relativistic Quantum
Theory}, Kluwer, Dordrecht (1993).}

[8] E.C.G. Stueckelberg, Helv. Phys. Acta {\bf
14}, 372, 585; {\bf 15}, 23 (1942).

[9] A. Einstein, {\it The Meaning of Relativity}, Princeton University Press, Princeton (1922).

[10] Isaac Newton, {\it Philosophia Naturalis
Principia Mathematica},
London (1687).

[11] I.B. Cohen and A. Whitman  {\it The
Principia: Mathematical Principles of Natural Philosophy: A  New
Translation}, University of California Press, Berkeley (1999).

[12] R.P. Feynman, Mathematical formulation of the quantum theory of electromagnetic interaction, Phys. Rev. {\bf 80}, 440 (1950).

[13] P.A.M. Dirac, {\it Quantum Mechanics}, 3rd edition, Oxford University Press, London (1947).

[14] A. Sommerfeld,{\it Atombau und Spektrallinean},vol. II, Chap. 4, Friedrich Vieweig and Sohn, Braunschweig (1939)

[15] J.L. Cook, Solution of the relativistic two-body problem II. Quantum mechanics, Australian Jour. Phys, {\bf 25},141 (1972)

[16] J.S. Zmuidzinas, Jour. Math. Phys. {\bf 7},
764 (1966).

{\obeylines
[17] P. Winternitz, personal communication,University of Montreal, Montreal, Quebec,Canada (1985).
[18] R.I. Arshansky and L.P. Horwitz, The relativistic two-body bound state, I. The spectrum, Jour. Math. Phys. {\bf 30}, 66 (1989).

[19] R.I. Arshansky and L.P.Horwitz, The relativistic two-body bound state, II. The induced representation of SL(2,C), Jour. Math. Phys. {\bf 30}, 380 (1989).

[20] M.A. Naimark, {\it Linear Representations of the Lorentz Group}, Pergamon Press, New York (1964).

[21] I.M. Gel'fand, R.A. Minlos and Z. Ya. Shapiro,
{\it Representations of the Rotation and Lorentz Groups and Their
Applications},Pergamon Press, New York (1963).

[22] V. Bargmann,  {\it Irreducible Unitary Representations of the Lorentz Group}, Annals of Mathematics {\bf 48}, No. 3 (1947).

[23] M. Bacry, personal communication, Tel Aviv University, Ramat Aviv, Israel (1990).

[24] A. Shapere and F. Wilczek, {\it Geometric Phases in Physics}, World Scientific, Singapore (1989).

[25] E. Merzbacher, {\it Quantum Mechanics}, 2nd edition,  Wiley, New York (1970).}

{\obeylines
[26]I.S. Gradshteyn and I.M. Rhyzik, {\it Table of Integrals, Series ad Products}, 7th edition, ed. Alan Jeffrey, D. Zwillinger, Academic Press, London (2007).

[27] E. Wigner, Ann. of Math.  {\bf 40} 149 (1939).

[28] A. Ashtekar, personal communication, University of Syracuse, Syracuse, New York (1982).

[29] G. Chew, {\it The Analytic S Matrix}, Benjamin, New York (1966).

[30] R. Hagedorn, I. Montvey, J. Rafelski, in {\it Hadronic Matter}, Erice 1978,p. 49,  eds. N. Cabbibo, L. Sertorio, Plenum, New York (1978)

[31] R. Hofstadter, R. Bumiller and M.R. Yearian, Electromagnetic structure of the proton and neutron, Rev. Mod. Phys. {\bf 30},482  (1958).

[32] C. Itzykson and J.-B. Zuber, {\it Quantum Field Theory}, McGraw-Hill, New York (1980).

[33] R.P. Feynman, M. Kislinger and F. Ravndal, Phys. Rev. {\bf D3}, 2706 (1971).}

{\obeylines
[34] Y.S. Kim and M.E. Noz, Covariant harmonic oscillators and excited baryon decays, Prog. Theor. Phys. {\bf  57} 1373 (1977).

[35] H. Leutwyler and J. Stern, Covariant quantum mechanics on a null plane, Phys. Lett. {\bf B 69 },207 (1977).}

{\obeylines
[36] Y. Dothan, M. Gell-Mann and Y. Ne'eman, Series of hadronic energy levels as representations of non-compact groups, Phys. Lett. {\bf 17}, 148 (1965).

[37] M. Land, Harmonic oscillator states with integer and non-integer orbital angular momentum,  Jour. of Phys. Conf. Series {\bf 330} 012014 (2011).

\end